\def\beg{\begin{equation}}
\def\eeq{\end{equation}}
\documentstyle[12pt]{article}
\begin{document}
\begin{center}
{\Large{\bf Comments on ``Anomalous-Filling-Factor-Dependent Nuclear Spin Polarization in a 2D Electron System: Quantum Hall Effect, by J. H. Smet, K. von Klitzing, et al, Phys. Rev. Lett. 92, 086802(2004)".}}
\vskip0.55cm
{\bf Keshav N. Shrivastava}
\vskip0.25cm
{\it School of Physics, University of Hyderabad,\\
Hyderabad  500046, India}
\end{center}
 We find that the nuclear spin polarization has not been treated correctly. The references given are those of wrong papers. The credits assigned for discoveries are also not correct. Incorrect theories have been cited. The reference to the correct theory has been neglected. Because, there are lots of good people, so they do not want to give reference to my papers even though they have not solved the problem of quantum Hall effect and we have.

\vskip1.0cm

\noindent {\bf 1.~ Introduction}

In a recent letter[1] published in the PRL, we find that our result is used without giving reference to our papers. Therefore, our comments are given below.\\

\noindent {\bf 2.~ Comments}

1. Page 086802-1, para 1. "The spin of the 2D electrons with density $n_{2D}$ has long been recognized ...".  Actually, the Laughlin's wave function does not depend on spin.
It requires only an odd number which makes the wave function antisymmetric. Halperin[2]
did suggest the possibility of singlet pairing at the half filled band. The spin singlet of two particles will give the charge of $2e$ as in superconductors not $e/2$. Therefore,
spin singlet is not the correct answer for the half filled band. Actually, there is a way to get $S=0$ and charge $e/2$ but it is not through Halperin's paper. It can be obtained from my formula only.\\

2. Page 086802-1, para 2.``This spin transition physics can be lucidly phrased ... invoking composite fermion (CF)...". We have informed the first author by posting our paper ( K. N. Shrivastava, cond-mat/0304269), at an earlier time that this CF picture is entirely wrong. Imagine that two flux quanta are attached to one electron and this is called CF. When this object is heated, two flux quanta come out without any current. So flux can be produced without producing current.  That is why this CF model is entirely incorrect.\\

``This phase transition leaves unambiguous fingerprints in the transport properties. Because the gap diminishes as level ($0\downarrow$) overtakes($1,\uparrow$)...". The statement is correct but the reference to the original paper has been eaten up and the credit has not been given. The correct spin configuration is given by me in 1985[3]. At that time von Klitzing's group did not know the spin orientations.\\

3. Page 086802-2. Caption of Fig.1. ``... evolution of the composite fermion Landau levels with density or $B$ field at fixed $\nu$=2/3 ...". The actual objects are not ``composite fermions". The authors have misinterpreted the data.\\

4.Page 086802-2, right hand side. `` ... filling dependent nuclear spin polarization develops". The result is correct but the reference to the original paper and hence the credit to the original author has not been given.\\

5. Page 086802-3, left hand side. ``Its origin is puzzling. Most of all, what is the impetus for such a large change in $<I>$ upon switching to a different $\nu_{rest}$".\\
The answer is given in my book[4].\\

6. Page 086802-3, right hand side, line 22. `` Skyrmion physics makes $<S>$ drop to small values". The Skyrmion is irrelevant to the present problem.\\
``Spin-polarized CF metallic state is anticipated". The CF is irrelevant to the quantum Hall effect.\\

7. Page 086802-4, left hand side: ``large filling factor dependence of the spin polarization of nuclei residing in the same plane as the 2D electrons has been revealed". The authors have reported as if it is a new result but such a result is given in my book[4] earlier.\\
\noindent{\bf 3.~~Addenda}

     Recently, it has been pointed out by Halperin that there is a Pfaffian state, which is defined by,
\beg
Pfaff\,\, of\,\, M_{i,j}= {1\over 2^{L/2}(L/2)!}\Sigma_{\sigma\epsilon S_L}sgn \sigma\Pi_{k=1}^{L/2}M_{\sigma(2k-1),\sigma(2k)}
\eeq
\beg
\psi_{Pfaff}=Pfaff\,\,\, of \,\,\,{1\over z_i-z_j}\psi_{Laughlin}.
\eeq
Therefore, Pfaffian is not a cure for Laughlin and hence serves no useful purpose what so ever. Here $M_{i,j}$ are the matrix elements of $L\times L$ antisymmetric matrix. $S_L$ is the mermutation group on L objects and $\psi_{Laughlin}$ is the Laughlin's wave function containing $(z_i-z_j)^q$ with q=odd number =3. $\psi_{Pfaff}$ is a wave function for spinless electrons in the lowest Landau level. The composite $\psi^+(z)U(z)^q$ is a neutral boson for q=odd and neutral fermion if q is even. So for q=odd, we can get Bose-Einstein condensation which is Laughlin's state. However, incompressibility has been introduced artificially. Usually c=1 is used to set the units but $a_o$=1 which gives incompressibility is algebraically not correct because the magnetic length is never unity.\\

There are lots of good people in Columbia, Princeton, Stanford, Stuttgart, Urbana, etc. and that is the explanation for not giving reference to my paper. However, these good people have not solved the problem of quantum Hall effect and they are not able to derive the 1/3 charge. Therefore, mine is the only correct theory[3]. Our results are being slowly pinched by the PRL authors without giving reference to our paper. Therefore, we are bound to make the observation that there is lack of desire to give the credit where it belongs.

\noindent{\bf4.~~ Conclusions}.

    The correct interpretation of the quantum Hall effect is given by us but Smet et al[1] while using spin combinations have not given the reference to our work. This is highly objectionable. The interpretation in terms of CF is entirely incorrect.
 \vskip1.0cm

\noindent{\bf4.~~References}
\vskip1.0cm
\begin{enumerate}
\item J. H. Smet, R. A. Deutschmann, F. Ertl, W. Wegschneider, G. Abstreiter and K. von Klitzing, Phys. Rev. Lett. {\bf 92}, 086802(2004).
\item B. I. Halperin, Helv. Phys. Acta {\bf 56}, 75 (1983)
\item K. N. Shrivastava, Phys. Lett. A {\bf 113}, 435 (1986)
\item K. N. Shrivastava, Introduction to quantum Hall effect,\\
      Nova Science Pub. Inc., N. Y. (2002).
\end{enumerate}

\end{document}